\documentclass[aps,amsmath,amssymb,twocolumn]{revtex4}

\usepackage{epsfig}
\usepackage{colordvi}
\usepackage{graphicx}
\usepackage{dcolumn}
\usepackage{bm}

\begin{document}

\title{Differential Dynamic  Microscopy of Bacterial Motility}

\author{L. G. Wilson}
\author{V. A. Martinez}
\author{J. Schwarz-Linek}
\author{J. Tailleur}
\author{P. N. Pusey}
\author{W. C. K. Poon}
\affiliation{SUPA and COSMIC, School of Physics \& Astronomy, The University of Edinburgh, Mayfield Road, Edinburgh EH9 3JZ, United Kingdom}

\author{G. Bryant}
\affiliation{Applied Physics, School of Applied Sciences, RMIT University, Melbourne, Victoria 3000, Australia}

\date{\today}

\begin{abstract}
We demonstrate `differential dynamic microscopy' (DDM) for the fast, high throughput characterization of the dynamics of active particles. Specifically, we characterize the swimming speed distribution and the fraction of motile cells in suspensions of {\em Escherichia coli} bacteria. By averaging over $\sim 10^4$ cells, our results are highly accurate compared to conventional tracking. The diffusivity of non-motile cells is enhanced by an amount proportional to the concentration of motile cells.  
\end{abstract}
\maketitle

Diverse processes in multicellular organisms such as chemotaxis involve motility \cite{Bray}, which is also ubiquitous in unicellular organisms such as bacteria, enabling, e.g. the pathogen {\em Helicobacter pylori} to invade the stomach epithelium \cite{MontecuccoRP_NatureMolCellBiol01}. Globally, bacterial motility may be coupled to aquatic nutrient recycling \cite{Azam01}. The bacterium {\em Escherichia coli} is a paradigm for understanding cell motility \cite{BergBook}. A cell executes a random walk by alternating between swimming (or `running') at average speed $\bar{v} \gtrsim 10 \mu$m/s for $\sim 1$~s and tumbling for $\sim 0.1$~s.  

Early bacterial motility work relied on tracking one to a few cells \cite{BergDB_Nature72,Schneider74}. Today, $\sim 10^2 - 10^3$ cells can be tracked simultaneously \cite{Worku98,WuJRSKDKMD_ApplEnvMicrobiol06,DouarcheABHSAL_PRL09}. Tracking yields a host of parameters, including $\bar{v}$ (e.g. in polymer solutions \cite{Schneider74}) and the fraction of motile organisms, $\alpha$ (e.g. in oceanic bacteria \cite{Azam01}). 
But tracking is laborious, and the need for averaging over many data sets to achieve high accuracy restricts the scope for time-dependent measurements. 

We demonstrate a fast, high throughput method for characterizing {\em E. coli} motility. It should be applicable to other bacteria and micro-organisms, and to a new generation of synthetic, self-propelled `active particles' \cite{Sen10}.

Dynamic light scattering (DLS), long used for measuring diffusivity in colloids, is in principle suitable for the fast characterization of motile bacteria \cite{NossalSCCL_OptComm71}. DLS yields the normalized intermediate scattering function (ISF), $f(q, \tau)$ (where $q$ is the scattering vector and $\tau$ is time) \cite{Berne00}, which probes density relaxation processes at length scale $2\pi/q$. But the lowest scattering angle in conventional DLS, $\sim 20^{\circ}$ (or $q \sim 4.5 \mu\mbox{m}^{-1}$), probes dynamics at $2\pi/q \lesssim 1.4\mu$m, where cell body precession \cite{BoonRNSC_BiophysJ74} and other motions in {\em E. coli} contribute strongly to the decay of the ISF. Thus, contrary to initial claims \cite{NossalSCCL_OptComm71}, {\em E. coli} swimming, which occurs on the scale of $\bar{v}/\tau_{\rm run} \sim 10\mu$m, cannot be characterized unambiguously using DLS unless we can access $q \lesssim 2\pi/10\mu{\rm m} \sim 0.6 \mu$m$^{-1}$ (or $\lesssim 3^{\circ}$) \cite{BoonRNSC_BiophysJ74}.

Instead of implementing such ultra-low-angle DLS, we use the powerful technique of Differential Dynamic Microscopy (DDM) to measure $f(q,\tau)$ for bacterial swimming. A form of DDM was first used to study density fluctuations in binary mixtures \cite{Giglio06}. It has recently been used to measure colloidal diffusivity \cite{CerbinoVT_PRL08}, requiring only non-specialized equipment (microscope, camera and computer). The DDM of colloids, however, does not utilize its unique capability to reach very low $q$ ($\lesssim 1\mu$m$^{-1}$), which turns out to be essential for probing bacterial swimming. 

The theory of DDM is detailed in \cite{GiavazziDBVTTBRC_PRE09}. We give an alternative derivation, which also explains experimental procedures. The raw data are time-lapsed images of (say) bacteria, described by the intensity $I(\vec{r},t)$ in the image plane ($\vec{r}$). From these we calculate difference images at various delay times, $\tau$, $D(\vec{r},\tau) = I(\vec{r},t + \tau) - I(\vec{r},t) = \Delta I(\vec{r},\tau) - \Delta I(\vec{r},0)$, where $\Delta I(\vec{r},t) = I(\vec{r},t) - \langle I \rangle$ denotes intensity fluctuations. Fourier transforming $D(\vec{r},\tau)$ gives
\begin{equation}
F_D (\vec{q},\tau)  =  \int D(\vec{r},\tau) e^{i\vec{q}\cdot \vec{r}} {\rm d}\vec{r}. \label{DICF}
\end{equation}
For stationary, isotropic processes, we average over the start time $t$ in the difference images and azimuthally in $\vec{q}$ space to calculate the basic output of DDM, what we may call the `differential intensity correlation function' (DICF), $\langle |F_D (q,\tau)|^2 \rangle$ (where $q = |\vec{q}|$).

We now show that the DICF is related simply to the ISF if we assume that intensity fluctuations in the image are proportional to the fluctuations in the number density of bacteria around the average density $\langle \rho \rangle$: 
\begin{equation}
\Delta I(\vec{r},t) = \kappa \Delta \rho(\vec{r},t)\;. \label{assumption}
\end{equation}
Here the constant $\kappa$ depends on the contrast mechanism and $\Delta \rho(\vec{r},t) = \rho(\vec{r},t) - \langle \rho \rangle$. Eqs.~(\ref{DICF}) and (\ref{assumption}) now give
\begin{eqnarray}
F_D (\vec{q},\tau) & = & \kappa [\Delta \rho (\vec{q},\tau ) - \Delta \rho (\vec{q},0)]\;,\\
\mbox{where}\;\; \Delta \rho (\vec{q},\tau ) & = & \int \Delta \rho (\vec{r},t)e^{i\vec{q}\cdot \vec{r}} {\rm d} \vec{r}\;. \label{deltarho}
\end{eqnarray}
Thus, the DICF can be expressed as
\begin{eqnarray}
\langle |F_D (q,\tau)|^2 \rangle & = & A(q) \left[ 1 - \frac{\langle \Delta \rho (q,0)\Delta \rho (q,\tau)\rangle}{\langle [\Delta \rho (q)]^2 \rangle}\right] \\ \label{FD1}
\mbox{where}\;\; A(q) & = & 2\kappa^2 \langle [  \Delta  \rho (q)]^2 \rangle\;.
\end{eqnarray}
The prefactor $A(q)$ depends on the imaging system, $\kappa$, and on the sample's structure, $\langle [ \Delta \rho (q)]^2 \rangle$. Recognizing that the $\tau$-dependent term on the right-hand side of Eq.~(\ref{FD1}) is the ISF, we arrive at this key result:
\begin{equation}
\langle |F_D (q,\tau)|^2 \rangle = A(q)\left[ 1 - f(q,\tau ) \right] + B(q)\;, \label{ddm}
\end{equation}
where we have included a term $B(q)$ to account for camera noise. Thus, the power spectrum of intensity fluctuations of the images, $\langle |F_D (q,\tau)|^2 \rangle$, yields the ISF. In practice, we reconstruct $f(q,\tau)$ by using a parametrized model of the ISF to fit the measured DICF with Eq.~(\ref{ddm}).

\begin{figure}[t]
\includegraphics*[width=3.2in]{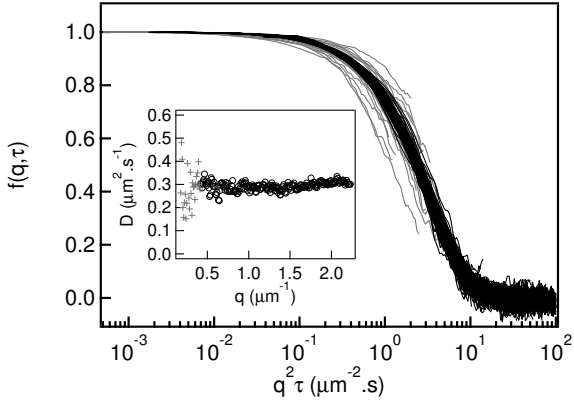}
\caption {Reconstructed ISFs of non-motile bacteria plotted against $q^2 \tau$. Solid (black) curves for over 200 values of $q$ in the range $0.5 \mu{\rm m}^{-1} \lesssim q \lesssim 2.2 \mu{\rm m}^{-1}$ collapse, but curves from lower $q$ (grey) do {\em not} collapse. Inset: fitted diffusivity $D(q)$ (black and grey with the same meaning).}\label{fig:motA}
\end{figure}

For independent particles, $f(q,\tau) = \langle e^{-i\vec{q}\cdot \Delta \vec{r}(\tau)}\rangle$, where $\Delta \vec{r}(\tau)$ is the single-particle displacement \cite{Berne00}. This reduces to $f(q,\tau) = e^{-Dq^2 \tau}$ for identical diffusing spheres with diffusivity $D$ \cite{Berne00}. For a swimmer with velocity $\vec{v}$, $\Delta \vec{r}(\tau) = \vec{v}\tau$. For an isotropic population of such swimmers in 3D, $f(q,\tau) = \text{sin} (qv\tau)/qv\tau \equiv \text{sinc} (qv\tau)$ \cite{Berne00}; since a swimmer inevitably also undergoes Brownian motion, this needs to be multiplied by an exponential prefactor $e^{-Dq^2 \tau}$. If only a fraction $\alpha$ of swimmers are motile with speed distribution $P(v)$, then the full ISF reads \cite{Stock_BiophysJ78}:
\begin{equation}
f(q,\tau) = e^{-Dq^2 \tau}\left[(1-\alpha) + \alpha \! \int_{0}^{\infty} \!\!\!\!\! P(v) \text{sinc} (qv\tau) {\rm d}v\right]. \label{master}
\end{equation}   
In order to use this model to interpret our DDM data from {\em E. coli}, we need to specify a form for $P(v)$. Limited previous data \cite{NossalSCCL_OptComm71,Stock_BiophysJ78} suggest a peaked function with $P(v \rightarrow 0) \rightarrow 0$. We use a Schulz distribution: 
\begin{equation}
P(v) = \frac{v^Z}{Z!}\left( \frac{Z+1}{\bar{v}} \right)^{Z+1} \exp \left[ - \frac{v}{\bar{v}}(Z+1) \right]\;, \label{schulz}
\end{equation}
where $Z$ is related to the variance $\sigma^2$ of the distribution by $\sigma = \bar{v}(Z+1)^{-1/2}$. The integral in Eq.~(\ref{master}) evaluates to \cite{pusey84}
\begin{eqnarray}
\int_{0}^{\infty} \!\!\!\!\!\!\! P(v)  \text{sinc}(qv\tau) {\rm d}v  & = & \left( \frac{Z+1}{Z q\bar{v}\tau} \right) \!
\frac{ \sin \left( Z \tan^{-1} \theta \right)}{\left(  1  + \theta^2 \right)^{Z/2} } \label{puseyA} \\
\mbox{where}\; \;  & \theta & =  (q\bar{v}\tau)/(Z+1). \label{puseyB}
\end{eqnarray}

We studied \emph{E. coli} AB1157 grown at 30$^{\circ}$C in L-broth, re-inoculated into T-broth and harvested in mid-exponential phase, washed three times by filtration (0.45$\mu$m filter) in motility buffer and re-suspended in the same buffer to an optical density of ~0.3 (at 600nm), giving a final cell volume fraction of $\phi \approx 0.06\%$. (See Supplementary Material for details.)  Care was taken throughout to minimize damage to flagella. A $\sim 400\mu$m deep flat glass cell  was filled with $\sim 150\mu$l of cell suspension, sealed, and observed at 22$\pm 1^\circ$C.  Swimming behavior was constant over a 15 minute period.  We also used a non-motile mutant with `paralyzed' flagella ({\em motA}).

\begin{figure}[t]
\includegraphics*[width=3.2in]{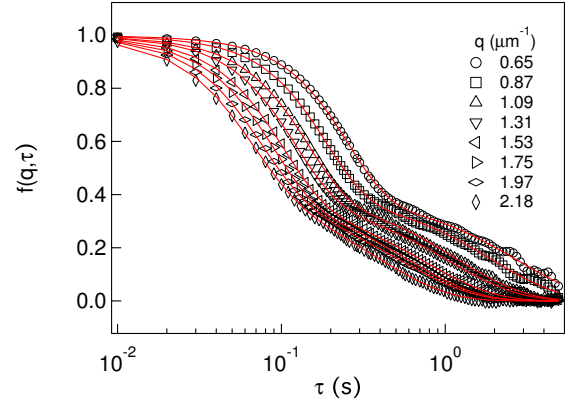}
\caption {Symbols: reconstructed ISFs for motile {\em E. coli} at 8 values of $q$ spanning the whole available range (see key). Lines: calculated ISFs, Eqs.~(\ref{master})-(\ref{puseyB}), using fitted parameters.}\label{fig:fits}
\end{figure}

We collected movies of cells using a 10$\times$ phase-contrast objective in a Nikon Eclipse Ti inverted microscope. Images were obtained $\approx 100\mu$m from the bottom of a $400 \mu$m-thick sample cell.  A high-speed camera (Mikrotron MC 1362) was connected to a PC with a frame grabber card with 1GB onboard memory. Movies were acquired typically at 100Hz. The frame size $L^2$ was $500 \times 500$ and $1024 \times 1024$ pixels for {\em motA} mutants and wild-type cells respectively, imaging $\sim 10^4$ cells in a 0.7mm$^2$ or 1.4mm$^2$ field of view over 38s or 8s. The pixel size (or spatial sampling frequency) is $k = 0.712\mu \text{m}^{-1}$, so that $q_{\rm min}=2\pi k/L \approx 0.01\mu \text{m}^{-1}$ or 0.004~$\mu$m$^{-1}$.  

To calculate the DICFs from the raw images, we used a LabView (National Instruments) code optimized for an 8-core PC (dual Intel Xeon quad-core processors, 2GHz/core, 4GB RAM). Analyzing $\sim 40$s of movies takes $\sim 10$~minutes.
We then fitted each DICF to Eq.~(\ref{ddm}) using Eqs.~(\ref{master})-(\ref{puseyB}). 
At each $q$, non-linear least-squares fitting using the Levenberg-Marquardt algorithm \cite{Recipes} in IGOR Pro (WaveMetrics) returns six parameters: $\bar{v}$, $\sigma$, $D$, $\alpha$, $A$ and $B$. Fitting the whole $q$ range takes $\sim 30$~s. From the fitted $A(q)$ and $B(q)$, we obtain the {\em reconstructed} ISF using the measured DICF and Eq.~(\ref{ddm}). We also obtain the {\em calculated} ISF by using the fitted $\{\bar{v}, \sigma, D, \alpha\}$ in Eqs.~(\ref{master})-(\ref{puseyB}).

We first studied non-motile ({\em motA}) cells. Measured DICFs are well fitted using Eq.~(\ref{ddm}) with $f(q,\tau) = e^{-Dq^2 \tau}$ (i.e. Eq.~(\ref{master}) with $\alpha = 0$) (Supp. Fig. 1). The fitted diffusivity, $D(q)$, was $q$ independent within experimental uncertainties in the range $0.5 \mu{\rm m}^{-1} \lesssim q \lesssim 2.2 \mu{\rm m}^{-1}$, Fig.~\ref{fig:motA} inset, and averaged to $D = 0.30 \pm 0.01 \mu$m$^2$/s. Conventional DLS (data not shown) gave an exponential $f(q,\tau)$ and $D = 0.32 \pm 0.02 \mu$m$^2$/s, agreeing with DDM. The reconstructed ISFs collapse onto each other in the range $0.5 \mu{\rm m}^{-1} \lesssim q \lesssim 2.2 \mu{\rm m}^{-1}$ when plotted against $q^2 \tau$ (black curves, Fig.~\ref{fig:motA}), i.e. the non-motile cells are purely diffusive. At $q \lesssim 0.5 \mu{\rm m}^{-1}$, $f(q,\tau)$ has not decayed to zero at the longest time probed in our experiments, so that fitting becomes less reliable because of the difficulty in estimating $A(q)$ (cf. Eq.~(\ref{ddm})). The reconstructed ISFs therefore do not collapse under $q^2 \tau$ scaling and $D(q)$ is noisy (grey curves, Fig.~\ref{fig:motA}; crosses, Fig.~\ref{fig:motA} inset). 

We next studied motile cells. The measured DICFs (Supp. Fig. 2) were again fitted to Eq.~(\ref{ddm}), now using the full $f(q,\tau)$ in Eq.~(\ref{master}) and a Schulz $P(v)$, Eqs.~(\ref{schulz})-(\ref{puseyB}). A selection of the reconstructed ISFs is shown in Fig.~\ref{fig:fits} (points), where we also superimpose the calculated ISFs (curves). The ISFs display a characteristic shape, especially at low $q$: a fast decay dominated by swimming followed by a slower decay dominated by diffusion. (Compare also the different time axes in Figs.~\ref{fig:motA} and \ref{fig:fits}.)

\begin{figure}[tb]
\includegraphics*[width=3.3in]{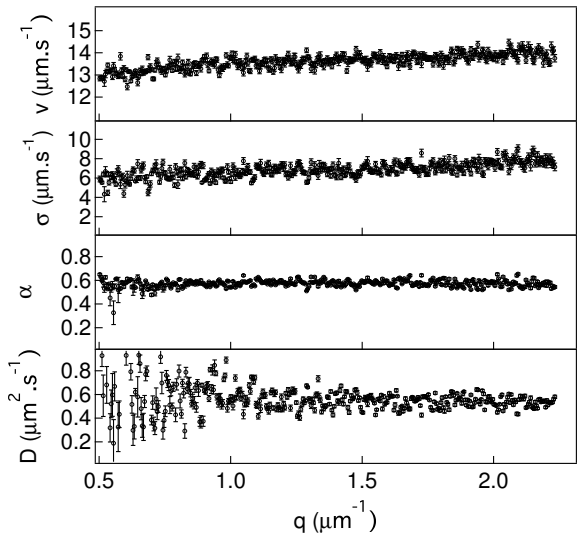}
\caption {Parameters extracted from fitting the DICF data for motile cells. From top to bottom: $\bar{v}$ and $\sigma$ of the Schulz distribution, motile fraction $\alpha$ and diffusivity $D$.}\label{fig:parms}
\end{figure}

All fit parameters characterizing swimming are shown in Fig.~\ref{fig:parms} \footnote{Using a log-normal distribution necessitated numerical integration in Eq.(\ref{master}) but did not change the conclusions.}. The noise increases at low $q$, primarily because the long-time, diffusive part of $f(q,\tau)$ has not reached zero in our time window at these $q$, Fig.~\ref{fig:fits}, rendering it harder to determine the diffusivity accurately: the low-$q$ noise is particularly evident in the fitted $D(q)$, Fig.~\ref{fig:parms}. But to within experimental uncertainties all parameters in Fig.~\ref{fig:parms} are essentially $q$-independent at least for $q \gtrsim 1 \mu$m$^{-1}$ \footnote{Fixing $D = D_0$, we obtained $q$ independent parameters only if $D_0$ was within $\pm 10\%$ of the freely-fitted value.}, suggesting that our model is able to capture essential aspects of the dynamics of a mixed population of non-motile and motile {\em E. coli}. Averaging yields  $\bar{v} = 13.7 \pm 0.1 \mu{\rm ms}^{-1}$ and $\bar{\sigma} = 7.0 \pm 0.1 \mu{\rm m s}^{-1}$, with error bars reflecting estimated residual $q$ dependencies. Changing $A(q)$ and $B(q)$ by using a $20\times$ objective (which is sub-optimal for our experiment) produced the same fitted motility parameters in the relevant $q$ range.

The presentation so far rests on a number of assumptions. Our derivation of Eq.~(\ref{ddm}) assumes that the decorrelation of $f(q,\tau)$ caused by the change in intensity of a swimmer's image due to its motion along the optic ($z$) axis can be neglected. While wild-type {\em E. coli} AB1157 tumbles between `runs' and the swim path between tumbles is slightly curved, Eq.~(\ref{master}) neglects these effects. We tested the validity of these assumptions by performing DDM analysis on the output from computer simulations.

\begin{figure} [t]
\includegraphics*[width=2.8in]{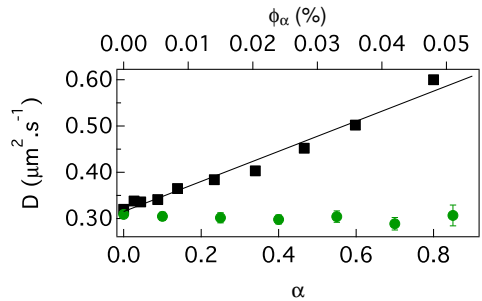}
\caption {$\blacksquare$: $D$ as a function of $\alpha$ (lower axis) and the volume fraction of swimmers, $\phi_{\alpha}$ (upper axis).  The black line is a linear fit. \OliveGreen{$\bullet$}: $D(\alpha)$ from simulations.}\label{fig:d_alpha}
\end{figure}

We carried out Brownian dynamics simulations of non-interacting point particles at a number density and in a geometry directly comparable to our experiments. A fraction $\alpha$ of the particles had a drift speed drawn from a Schulz distribution. From these simulations, we constructed a sequence of 2D pixellated `images' with the same field of view as in experiments. All particles in a slice of thickness $d$ centered at $z = 0$ contribute to the image. A particle at $(x,y,z)$ is `smeared' into an `image' covering the pixel containing $(x,y)$ and its 8 neighboring pixels. The contrast of the image, $c$, depends on $z$. We experimentally determined $d$ and $c(z)$by imaging a single bacterium as the focal plane traversed the cell. The measured $c(z)$ could be fitted by a symmetric quadratic that dropped to background noise outside a $\approx 40 \mu$m slice. 

As input, we used $\bar{v} = 13.7 \mu\mbox{ms}^{-1}$, $\sigma = 7.0 \mu\mbox{ms}^{-1}$, $\alpha = 0.577$ and $D = 0.543 \mu$m$^{2}$/s (cf. Fig.~\ref{fig:parms}). Fitting  DICFs calculated from simulated `images' (Supp. Fig. 3a) gave $q$-independent outputs (Supp. Fig. 3b): $\bar{v} = 13.8 \pm 0.1$, $\bar{\sigma} = 7.2 \pm 0.2$, $\bar{\alpha} = 0.58 \pm 0.01$ and $\bar{D} = 0.55 \pm 0.02$ (where the uncertainties are standard deviations), agreeing with inputs. Thus, at $d = 40 \mu$m depth of field, the intensity decorrelation due to $z$ motion has negligible effect, presumably because it is much slower than the decorrelation due to swimming and diffusion. However, if we scale $c(z)$ to smaller depths of field, the fitting eventually fails at $d \approx 10 \mu$m (data not shown): at this small focus depth, a small $z$ movement  produces a large intensity variation, invalidating our analysis.

DDM essentially determines the (inverse) time it takes a cell to traverse $\sim 2\pi/q$, i.e. it measures `linear speeds'. Tumbling or curvature will therefore lower the measured speed, with the effect more noticeable at lower $q$. Our experimental $v(q)$, Fig.~\ref{fig:parms}, does indeed show a slight decrease towards low $q$. As expected, however, the $v(q)$ recovered from analysing simulated straight swimmers (Supp. Fig. 3) show no such dependence. More detailed analysis of the measured $v(q)$ may therefore yield further information about tumbling and curvature. 

We next mixed suspensions of bacteria with known $\alpha$ with various proportions of non-motile cells, creating samples with $0 \leq \alpha \leq 0.8$. DDM shows that $D$ increases with $\alpha$, Fig.~\ref{fig:d_alpha} (squares). Since the fitting of $D$ from Eq.~\ref{master} is largely determined by the diffusion of the non-swimmers, Fig.~\ref{fig:d_alpha} shows that swimmers enhance the diffusion of non-swimmers. Note that this observation is {\em not} a fitting artifact: such dependence is {\em not} observed in simulations, which returned an $\alpha$-independent $D$, Fig.~\ref{fig:d_alpha}. Since the simulated particles are non-interacting, our experimental observation must be due to direct or hydrodynamic interaction between swimmers and non-swimmers. 

The enhanced diffusion of (passive) particles in suspensions of motile {\em E. coli} has been observed before using direct tracking at both low concentration ($\phi = 0.003\%$) in 3D \cite{Lubensky07} and high concentration ($\phi \approx 10\%$) in 2D \cite{Wu00}. Scaling arguments suggest that in the limit of independent swimmers, the enhancement should scale linearly as the concentration of swimmers $\phi_{\alpha}$ \cite{Graham08}. In our experiments, the volume fraction of non-swimmers varies, but remains $\lesssim 0.1\%$, i.e., they can be considered as independent `tracer' particles. Thus, Fig.~\ref{fig:d_alpha} can be reinterpreted as a plot of the effective diffusion coefficient of tracer particles as the concentration of swimmers increases from $\phi_{\alpha} = 0$ to $\phi_{\alpha} = 0.06\% \times 0.8 = 0.048\%$. The enhancement indeed scales linearly with the swimmer concentration. 

Our $D(\alpha)$ results may also be compared to enhanced tracer diffusion by {\em Chlamydomonas reinhardtii}, a nearly-spherical single-cell algae \cite{Goldstein09}. However, {\em C. reinhardtii} (radius $\sim 5 \mu$m) are larger than {\em E. coli}, and swim much faster ($\bar{v} \sim 100~\mu$ms$^{-1}$). More fundamentally, {\em C. reinhardtii} (a `puller') and {\em E. coli} (a `pusher') generate qualitatively different flow fields, which may have consequences for tracer diffusion \cite{Graham08}. Nevertheless, it is intriguing that 2\% of {\em C. reinhardtii} quadruples the diffusivity of $2 \mu$m tracers, while $0.048\%$ of motile {\em E. coli} already doubles the diffusivity of non-motile cells. 



To summarize, we have shown that DDM is a fast, high-throughput method for characterizing the bulk motility of wild-type {\em E. coli}. The method could, in principle, be extended to characterize the run-tumble-run random walk of individual cells \cite{AdlerMD_JGenMicrobiol67} (by going to even lower $q$), or to the study of motility near surfaces (which requires the use of a different $f(q,\tau)$ in Eq.~(\ref{master})).  The method may also be applicable to the study of other motile organisms, including spermatozoa, as well as for characterizing the motions of synthetic motile colloids \cite{Sen10}. But the $q$ range, camera speed and data acquisition time need to be optimized for each particular class of motility to be characterized. Our finding that even low concentrations of motile cells enhance the diffusivity of non-motile cells may have implications for understanding the coupling between bacterial motility and the recycling of organic debris in natural aqueous habitats \cite{Azam01}.  

Finally, we should emphasize that DDM yields $f(q,\tau)$ of suspensions of active swimmers irrespective of $\phi$, provided that Eq.~(\ref{assumption}) remains valid.  It is therefore a general method for studying the dynamics of these suspensions, including interaction effects at higher $\phi$, although new models will clearly be needed for interpreting the data.

{\em Acknowledgment} The work and part of GB's visit were funded by the EPSRC (EP/E030173 and EP/D071070).

\end{document}